\documentclass{llncs}

\usepackage{amsmath}
\usepackage{amsfonts}
\usepackage{amssymb}
\usepackage{graphicx}
\usepackage{todonotes}
\usepackage{boxedminipage}
\usepackage{bm}
\usepackage{subfigure}
\usepackage{cleveref}
\usepackage{paralist}

\Crefname{property}{Property}{Properties}

\title{Testing Upward Planarity of Partial $2$-Trees}

\author{Steven Chaplick\inst{1}\orcidID{0000-0003-3501-4608} 
	\and Emilio Di~Giacomo\inst{2}\orcidID{0000-0002-9794-1928} 
	\and Fabrizio Frati\inst{3}\orcidID{0000-0001-5987-8713} 
	\and Robert Ganian\inst{4}\orcidID{0000-0002-7762-8045} 
	\and Chrysanthi N. Raftopoulou\inst{5}\orcidID{0000-0001-6457-516X} 
	\and Kirill Simonov\inst{4}
}

\authorrunning{Chaplick et al.}

\institute{Maastricht University, Maastricht, The Netherlands,
	\email{s.chaplick@maastrichtuniversity.nl}
	\and Università degli Studi di Perugia, Perugia, Italy,
	\email{emilio.digiacomo@unipg.it}
	\and Roma Tre University, Rome, Italy, 	\email{frati@dia.uniroma3.it}
	\and Technische Universität Wien, Wien, Austria,
	\email{rganian@ac.tuwien.ac.at}
	\and National Technical University of Athens, Athens, Greece,
	\email{crisraft@mail.ntua.gr}
	}

\newcommand{\bigoh}{\ensuremath{\mathcal{O}}}

\newcommand{\NP}{\ensuremath{\mathrm{NP}}}

\renewcommand{\textrm}{\texttt}

\newcommand{\nullo}{\textsc{null}}

\newcommand{\shapeDesc}[8]{\ensuremath{\langle #1,\allowbreak #2,\allowbreak #3,\allowbreak #4,\allowbreak #5,\allowbreak #6,\allowbreak #7,\allowbreak #8 \rangle}}

\Crefname{figure}{Fig.}{Figs.}

\begin{document}

%\begin{CCSXML}
%<ccs2012>
%   <concept>
%       <concept_id>10003752.10003809.10010052</concept_id>
%       <concept_desc>Theory of computation~Parameterized complexity and exact algorithms</concept_desc>
%       <concept_significance>500</concept_significance>
%       </concept>
%   <concept>
%       <concept_id>10003752.10010061.10010063</concept_id>
%       <concept_desc>Theory of computation~Computational geometry</concept_desc>
%       <concept_significance>500</concept_significance>
%       </concept>
% </ccs2012>
%\end{CCSXML}
%
%\ccsdesc[500]{Theory of computation~Parameterized complexity and exact algorithms}
%\ccsdesc[500]{Theory of computation~Computational geometry}

	\maketitle

	\begin{abstract}
	We present an $O(n^2)$-time algorithm to test whether an $n$-vertex directed partial $2$-tree is upward planar. This result improves upon the previously best known algorithm, which runs in $O(n^4)$ time. 
	\end{abstract}
%
%\todo[inline,color=yellow]{FF: Cut .5 line from the abstract, change it on Easychair}
%\todo[inline,color=yellow]{FF: the command \ textcolor does not go well with line break for shape sequences. Anybody knows how to fix that?}

\section{Introduction}\label{sec:intro}

A digraph is \emph{upward planar} if it admits a drawing that is at the same time \emph{planar}, i.e., it has no crossings, and \emph{upward}, i.e., all edges are drawn as curves monotonically increasing in the vertical direction. Upward planarity is a natural variant of planarity for directed graphs and finds applications in those domains where one wants to visualize networks with a hierarchical structure. 

Upward planarity is a classical research topic in Graph Drawing since the early 90s. Garg and Tamassia have shown that recognizing upward planar digraphs is \NP-complete~\cite{gt-ccu-01}, however polynomial-time algorithms have been proposed for various cases, including digraphs with fixed embedding~\cite{bdl-udtd-94}, single-source digraphs~\cite{DBLP:journals/siamcomp/BertolazziBMT98,BrucknerHR19,HuttonL91,HuttonL96}, outerplanar digraphs~\cite{Papakostas94}. The case of directed partial $2$-trees, which is of central interest to this paper and includes, among others, series-parallel digraphs, has been investigated by Didimo et al.~\cite{DBLP:journals/siamdm/DidimoGL09} who presented an $\bigoh(n^4)$-time testing algorithm. The parameterized complexity of the upward planarity testing problem has also been investigated~\cite{Chan04,cdf-up-22,DBLP:journals/siamdm/DidimoGL09,HealyL06}. 

In this paper, we present an $\bigoh(n^2)$-time algorithm to test upward planarity of directed partial $2$-trees, improving upon the $\bigoh(n^4)$-time algorithm by Didimo et al.~\cite{DBLP:journals/siamdm/DidimoGL09}. There are two main ingredients that allow us to achieve such result. 

\begin{figure}[tbp]
	\centering
	\subfigure[]{\label{fig:expansionA}\includegraphics[scale=0.7, page=1]{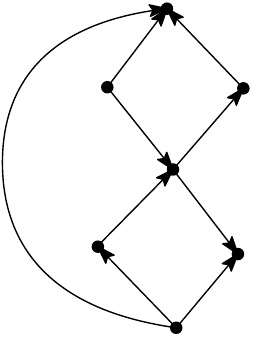}}
	\hfil
	\subfigure[]{\label{fig:expansionB}\includegraphics[scale=0.7, page=2]{figuresSP/Expansion}}
	\caption{\label{fig:expansion}Splitting a vertex in a non-expanded directed partial $2$-tree (a) might result in an expanded digraph (b) which is not a directed partial $2$-tree.}
\end{figure}

First, following the approach in~\cite{cdf-up-22}, our algorithm traverses the SPQ-tree of the input digraph $G$ while computing, for each component of $G$, the possible ``shapes'' of its upward planar embeddings. The algorithm in~\cite{cdf-up-22} only works for \emph{expanded} digraphs, i.e., digraphs such that every vertex has at most one incoming or outgoing edge. Although every digraph can be made expanded while preserving its upward planarity by ``splitting'' its vertices~\cite{DBLP:journals/siamcomp/BertolazziBMT98}, this modification might not maintain that the digraph is a directed partial $2$-tree; see Fig.~\ref{fig:expansion}. We present a novel algorithm that is applicable to non-expanded digraphs. We propose a new strategy to process P-nodes, which is simpler than the one of~\cite{cdf-up-22} and allows us to compute some additional information that is needed by the second ingredient. Further, we give a more efficient procedure than the one of~\cite{cdf-up-22} to process the S-nodes; this is vital for the overall running time of our algorithm.

Second, the traversal of the SPQ-tree $T$ of $G$ tests the upward planarity of $G$ with the constraint that the edge corresponding to the root of $T$ is incident to the outer face. Then $\bigoh(n)$ traversals with different choices for the root of $T$ can be used to test the upward planarity of $G$ without that constraint. However, following a recently developed strategy~\cite{dlop-ood-20,DBLP:journals/comgeo/Frati22}, in the first traversal of $T$ we compute some information additional to the possible shapes of the upward planar embeddings of the components of $G$. A clever use of this information allows us to handle P-nodes more efficiently in later traversals. Our testing algorithms can be enhanced to output an upward planar drawing, if one exists, although we do not describe the process explicitly. 

\paragraph*{Paper organization} In \Cref{sec:prelim} we give some preliminaries. In \Cref{sec:prescribed} we describe the algorithm for biconnected digraphs with a prescribed edge on the outer face, while in \Cref{sec:unrooting} we deal with general biconnected digraphs. \Cref{se:simply-connected} extends our result to simply connected digraphs. Future research directions are presented in \Cref{sec:conclusions}. Lemmas and theorems whose proofs are omitted are marked with a $(\star)$ and can be found in the full version of the paper.

\section{Preliminaries}\label{sec:prelim}

In a digraph, a \emph{switch} is a source or a sink. The \emph{underlying graph} of a digraph is the undirected graph obtained by ignoring the edge directions. When we mention connectivity of a digraph, we mean the connectivity of its underlying graph.

%A \emph{drawing} of a graph maps each vertex to a point in the plane and each edge to a Jordan arc between the points representing the end-vertices of the edge. A drawing is \emph{planar} if no two edges intersect, except at common end-vertices. A planar drawing partitions the plane into connected regions, called \emph{faces}. The unbounded face is the \emph{outer face}, while the other faces are \emph{internal}. 
A \emph{planar embedding} of a connected graph is an equivalence class of planar drawings, where two drawings are equivalent if: (i) the clockwise order of the edges incident to each vertex is the same; and (ii) the sequence of vertices and edges along the boundary of the outer face is the same. 

%is a triple $(e_1,u,e_2)$, where $e_1$ and $e_2$ are two edges that are consecutive in the clockwise order of the edges incident to a vertex $u$ and $f$ is the face that lies to the left of $e_1$ when traversing such an edge towards $u$.

\begin{figure}[tbp]
	\centering
	\subfigure[]{\label{fig:preliminariesA}\includegraphics[scale=0.7, page=1]{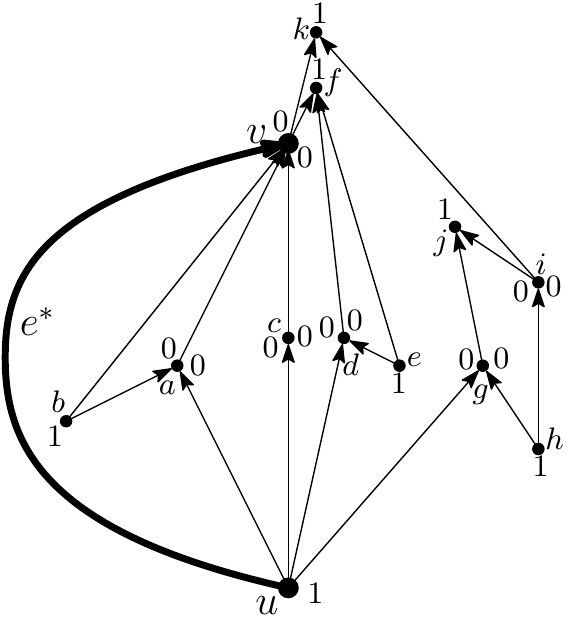}}
	\hfil
	\subfigure[]{\label{fig:preliminariesB}\includegraphics[scale=0.7, page=2]{figuresSP/Preliminaries}}
	\caption{\label{fig:preliminaries}(a) Labels at the angles of an upward planar embedding $\mathcal E$ of a biconnected directed partial $2$-tree $G$; the missing labels are $-1$. (b) An  SPQ-tree $T$ of $G$ with respect to $e^*$. The restriction of $\mathcal E$ to $G_{\sigma^*}$ is a $uv$-external upward planar embedding $\mathcal E_{\sigma^*}$ with shape description $\shapeDesc{0}{1}{1}{0}{\textrm{out}}{\textrm{out}}{\textrm{in}}{\textrm{out}}$. The shape sequence of $\mathcal E_{\sigma^*}$ is $[\shapeDesc{0}{0}{1}{1}{\textrm{out}}{\textrm{out}}{\textrm{in}}{\textrm{in}},
		\shapeDesc{0}{0}{1}{1}{\textrm{out}}{\textrm{out}}{\textrm{in}}{\textrm{in}},
		\shapeDesc{-1}{1}{1}{1}{\textrm{out}}{\textrm{out}}{\textrm{out}}{\textrm{out}},
		\shapeDesc{-1}{1}{1}{1}{\textrm{out}}{\textrm{out}}{\textrm{out}}{\textrm{out}}]$. The contracted shape sequence of $\mathcal E_{\sigma^*}$ is $[\shapeDesc{0}{0}{1}{1}{\textrm{out}}{\textrm{out}}{\textrm{in}}{\textrm{in}},
		\shapeDesc{-1}{1}{1}{1}{\textrm{out}}{\textrm{out}}{\textrm{out}}{\textrm{out}}]$.}
\end{figure}

A drawing of a digraph is \emph{upward} if every edge is represented by a Jordan arc whose $y$-coordinates monotonically increase from the source to the sink of the edge. A drawing of a digraph is \emph{upward planar} if it is both upward and planar. An upward planar drawing of a graph determines an assignment of labels to the angles of the corresponding planar embedding, where an \emph{angle} $\alpha$ at a vertex $u$ in a face $f$ of a planar embedding represents an incidence of $u$ on $f$. Specifically, $\alpha$ is \emph{flat} and gets label $0$ if the edges delimiting it are one incoming and one outgoing at $u$. Otherwise, $\alpha$ is a \emph{switch} angle; in this case, $\alpha$ is \emph{small} (and gets label $-1$) or \emph{large} (and gets label $1$) depending on whether the (geometric) angle at $f$ representing $\alpha$ is smaller or larger than $180^\circ$, respectively, see Fig.~\ref{fig:preliminariesA}. An \emph{upward planar embedding} is an equivalence class of upward planar drawings of a digraph $G$, where two drawings are equivalent if they determine  the same planar embedding $\mathcal E$ for $G$ and  the same label assignment for the angles of $\mathcal E$. 
%The label assignments that enhance a planar embedding to an upward planar embedding have been characterized as follows. 

\begin{theorem}[\cite{bdl-udtd-94,DBLP:journals/siamdm/DidimoGL09}]\label{th:upward-conditions}
	Let $G$ be a digraph with planar embedding  $\mathcal E$, and $\lambda$ be a label assignment for the angles of $\mathcal E$. Then $\mathcal E$ and $\lambda$ define an upward planar embedding of $G$ if and only if the following hold:
	\begin{enumerate}[]
		\item{\bf (UP0)} If $\alpha$ is a switch angle then $\alpha$ is small or large, otherwise it is flat.
		\item{\bf (UP1)} If $v$ is a switch vertex, the number of small, flat and large angles incident to $v$ is equal to $\deg(v)-1$, $0$, and $1$, respectively.
		\item{\bf (UP2)} If $v$ is a non-switch vertex, the number of small, flat and large angles incident to $v$ is equal to $\deg(v)-2$, $2$, and $0$, respectively.
		\item{\bf (UP3)} If $f$ is an internal face (the outer face) of $\mathcal E$, the number of small angles in $f$ is equal to the number of large angles in $f$ plus $2$ (resp.\ minus $2$).
	\end{enumerate}
\end{theorem}

%%%%%%%%%%%%%%%%%%%%%%%%%%%%%%%%%%%%%%%%%%%
%%%%%%%%%%%%%%%%%%%%%%%%%%%%%%%%%%%%%%%%%%%
%%%%%%%%%%%%%%%%%%%%%%%%%%%%%%%%%%%%%%%%%%%

%Given an undirected graph $H$ and two vertices $s$ and $t$ of $H$, we say that $H$ is \emph{$st$-biconnectible} if it is $2$-connected or if it becomes so by adding the edge $(s,t)$ to it. 

The class of \emph{partial $2$-trees} can be defined equivalently as the graphs with treewidth at most two, or as the graphs that exclude $K_4$ as a minor, or as the subgraphs of the $2$-trees. Notably, it includes the class of \emph{series-parallel graphs}. 

\begin{figure}[tbp]
	\centering
	\subfigure[]{\label{fig:preliminariesC}\includegraphics[scale=0.7, page=3]{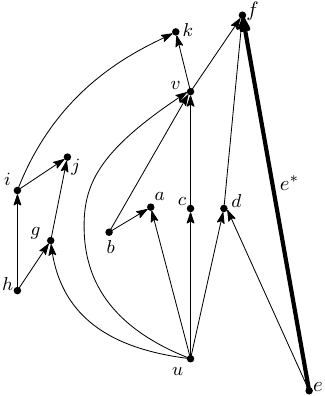}}
	\hfil
	\subfigure[]{\label{fig:preliminariesD}\includegraphics[scale=0.7, page=4]{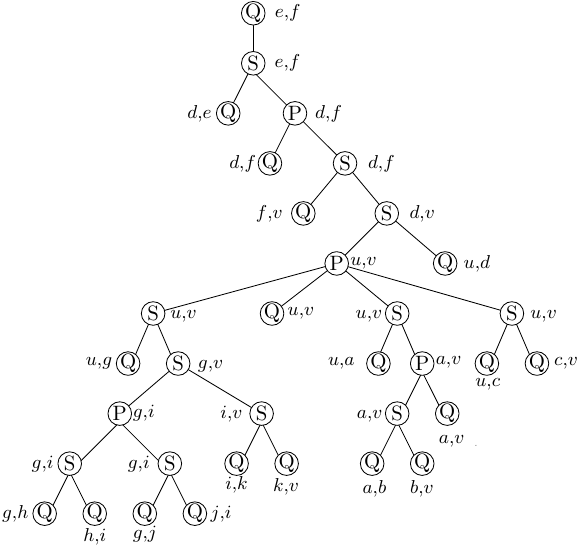}}\\
	\subfigure[]{\label{fig:preliminariesE}\includegraphics[scale=0.7, page=5]{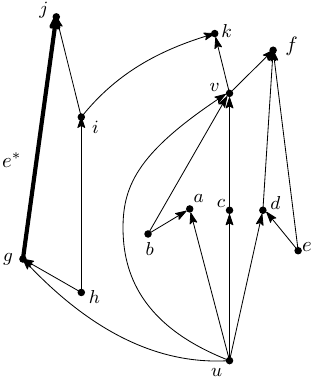}}
	\hfil
	\subfigure[]{\label{fig:preliminariesF}\includegraphics[scale=0.7, page=6]{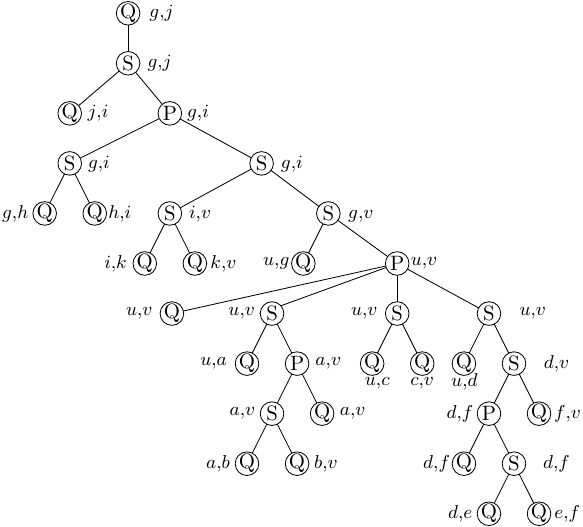}} 
	\caption{\label{fig:preliminariesCF} Two different choices for the root of the SPQ-tree of \Cref{fig:preliminaries}. The reference edge is shown in bold.}

\end{figure}

Let $G$ be a biconnected partial $2$-tree and let $e^*$ be an edge of $G$. An \emph{SPQ-tree $T$ of $G$ with respect to $e^*$} (see Fig.~\ref{fig:preliminariesB}) is a tree that describes a recursive decomposition of $G$ into its ``components''. SPQ-trees are a specialization of \emph{SPQR-trees}~\cite{dt-opl-96,gm-lti-00}. Each node $\mu$ of $T$ represents a subgraph $G_{\mu}$  of $G$, called the \emph{pertinent graph} of $\mu$, and is associated with two special vertices of $G_{\mu}$, called \emph{poles} of $\mu$. The nodes of $T$ are of three types: a Q-node $\mu$ represents an edge whose end-vertices are the poles of $\mu$, an S-node $\mu$ with children $\nu_1$ and $\nu_2$  represents a series composition in which the components $G_{\nu_1}$ and $G_{\nu_2}$ share a pole to form $G_{\mu}$, and a P-node $\mu$ with children $\nu_1, \dots, \nu_k$ represents a parallel composition in which the components $G_{\nu_1}, \dots , G_{\nu_k}$ share both poles to form $G_{\mu}$. The root of $T$ is the Q-node representing the edge $e^*$. %Details are in the appendix.
By our definition, every S-node has exactly two children that can also be S-nodes; because of this assumption, the SPQ-tree of a biconnected partial $2$-tree is not unique. However, from an SPQ-tree $T$, we can obtain an SPQ-tree of $G$ with respect to another reference edge $e^{**}$ by selecting the Q-node representing $e^{**}$ as the new root of $T$ (see \Cref{fig:preliminariesCF}).

%Also, the SPQ-tree is not unique in general, because of the different  choices of the cut-vertex $w$ for S-nodes. Similarly, selecting another reference edge, one may construct a different SPQ-tree. In order to have a ``unique" SPQ-tree for all reference edges, we equip nodes with \emph{labels} as follows. The label for P-nodes consists of its two poles, while for S-nodes the label contains the two poles and the cut-vertex used for computing its two children.  Assume that we have computed the SPQ-tree with respect to a reference edge $e^*$, and let $(u,v,w)$ be a label of an S-node.
%When computing the decomposition from a different edge $e^{**}$ and when creating an S-node whose poles are two among $u$, $v$, and $w$ and whose pertinent graph contains all of $u$, $v$, and $w$, we select as cut-vertex the unique non-pole  among $u$, $v$, and $w$, i.e. the S-node is labeled as $(u,v,w)$. As stated in the following lemma, this process gives the same labeled tree $T$ for every reference edge. Thus, we can talk about the (unrooted) SPQ-tree $T$ of a biconnected partial $2$-tree and define its root independently (the edge $e^*$ that defines the initial decomposition is chosen in any way). 

A \emph{directed partial $2$-tree} is a digraph whose underlying graph is a partial $2$-tree. When talking about an SPQ-tree $T$ of a biconnected \emph{directed} partial $2$-tree $G$, we always refer to an SPQ-tree of its underlying graph, although the edges of the pertinent graph of each node of $T$ are oriented as in $G$. Let $\mu$ be a node of $T$ with poles $u$ and $v$. A \emph{$uv$-external upward planar embedding} of $G_{\mu}$ is an upward planar embedding of $G_{\mu}$ in which $u$ and $v$ are incident to the outer face. In our algorithms, when testing the upward planarity of $G$, choosing an edge $e^*$ of $G$ as the root of $T$ corresponds to requiring $e^*$ to be incident to the outer face of the sought upward planar embedding $\mathcal E$ of $G$. For each node $\mu$ of $T$ with poles $u$ and $v$, the restriction of $\mathcal E$ to $G_{\mu}$ is a $uv$-external upward planar embedding $\mathcal E_{\mu}$ of $G_{\mu}$. In~\cite{cdf-up-22}, the possible ``shapes'' of the cycle bounding the outer face $f_{\mu}$ of $\mathcal E_{\mu}$ have been described by the concept of \emph{shape description}. This is the tuple $\shapeDesc{\tau_l}{\tau_r}{\lambda(u)}{\lambda(v)}{\rho^l_u}{\rho^r_u}{\rho^l_v}{\rho^r_v}$, defined as follows. Let the \emph{left outer path} $P_l$ (the \emph{right outer path} $P_r$) of $\mathcal E_{\mu}$ be the path that is traversed when walking from $u$ to $v$ in clockwise (resp.\ counterclockwise) direction along the boundary of $f_{\mu}$. The value $\tau_l$, called \emph{left-turn-number} of $\mathcal E_{\mu}$, is the sum of the labels of the angles at the vertices of $P_l$ different from $u$ and $v$ in $f_{\mu}$; the \emph{right-turn-number} $\tau_r$ of $\mathcal E_{\mu}$ is defined similarly. The values $\lambda(u)$ and $\lambda(v)$ are the labels of the angles at $u$ and $v$ in $f_{\mu}$, respectively. The value $\rho^l_u$ is $\textrm{in}$ ($\textrm{out}$)
if the edge incident to $u$ in $P_l$ is incoming (outgoing) at $u$; the values $\rho^r_u$, $\rho^l_v$, and $\rho^r_v$ are defined similarly. The values of a shape description depend on each other, as in the following.

\begin{observation}[\cite{cdf-up-22}]\label{obs:dependence-parameters}
	The shape description $\shapeDesc{\tau_l}{\tau_r}{\lambda(u)}{\lambda(v)}{\rho^l_u}{\rho^r_u}{\rho^l_v}{\rho^r_v}$ of $\mathcal E_{\mu}$ satisfies the following properties:
	\begin{enumerate}[(i)]
		\item $\rho^l_u$ and $\rho^r_u$ have the same value if $\lambda(u)\in \{-1,1\}$, while they have different values if $\lambda(u)=0$;
		\item $\rho^l_v$ and $\rho^r_v$ have the same value if $\lambda(v)\in \{-1,1\}$, while they have different values if $\lambda(v)=0$;
		\item $\rho^l_u$ and $\rho^l_v$ have the same value if $\tau_l$ is odd, while they have different values if $\tau_l$ is even;
		\item $\tau_l+\tau_r+\lambda(u)+\lambda(v)=2$.
	\end{enumerate}
\end{observation}

A set $\mathcal S$ of shape descriptions  is \emph{$n$-universal} if, for every $n$-vertex biconnected directed partial $2$-tree $G$, for every rooted SPQ-tree $T$ of $G$, for every node $\mu$ of $T$ with poles $u$ and $v$, and for every $uv$-external upward planar embedding $\mathcal E_{\mu}$ of $G_{\mu}$, the shape description of $\mathcal E_{\mu}$ belongs to $\mathcal S$. Thus, an $n$-universal set is a super-set of the \emph{feasible set $\mathcal F_{\mu}$ of $\mu$}, that is, the set of shape descriptions $s$ such that $G_{\mu}$ admits a $uv$-external upward planar embedding with shape description~$s$. Our algorithm will determine $\mathcal F_{\mu}$ by inspecting each shape description $s$ in an $n$-universal set and deciding whether $G_{\mu}$ admits a $uv$-external upward planar embedding with shape description $s$ or not. We have the following lemmas.

\begin{lemma}[$\star$]\label{le:shape-generation}
An $n$-universal set $\mathcal S$ of shape descriptions with $|\mathcal S|\in \bigoh(n)$ can be constructed in $\bigoh(n)$ time. 
\end{lemma}

\begin{lemma}[$\star$]\label{le:shape-data-structure}
Any subset $\mathcal F$ of an $n$-universal set can be stored in $\bigoh(n)$ time and space and querying whether a shape description is in $\mathcal F$ takes $\bigoh(1)$ time.
\end{lemma}

%\begin{proof}
%From the proof of Lemma~\ref{le:shape-generation}, we have that the left-turn-number of any shape description in $\mathcal F$ is in the range $[-n,n]$ and that there are $\bigoh(1)$ shape descriptions in $\mathcal F$ whose left-turn-number is the same. We can thus store $\mathcal F$ in $\bigoh(n)$ time and space in an array with $2n+1$ elements where, for $i=0,\dots,2n$, the element with index $i$ stores all the shape descriptions in $\mathcal F$ whose left-turn-number is $i-n$. A query on whether a shape description $s$ belongs to $\mathcal F$ can be answered by accessing in $\bigoh(1)$ time the element of the array with index $\tau+n$, where $\tau$ is the value of the first element of $s$, and then searching in $\bigoh(1)$ time whether $s$ coincides with any of the $\bigoh(1)$ shape descriptions stored at that element. 
%\end{proof}

Consider a P-node $\mu$ in an SPQ-tree $T$ of a biconnected directed partial $2$-tree $G$. Let $\nu_1,\dots,\nu_k$ be the children of $\mu$ in $T$. Consider any $uv$-external upward planar embedding $\mathcal E_{\mu}$ of $G_{\mu}$. For $i=1,\dots,k$, the restriction of $\mathcal E_{\mu}$ to $G_{\nu_i}$ is a $uv$-external upward planar embedding $\mathcal E_{\nu_i}$ of $G_{\nu_i}$; let $\sigma_i$ be the shape description of $\mathcal E_{\nu_i}$. Assume  that $\mathcal E_{\nu_1},\dots,\mathcal E_{\nu_k}$ appear in this clockwise order around $u$, where the left outer path of $\mathcal E_{\nu_1}$ and the right outer path of $\mathcal E_{\nu_k}$ delimit the outer face of $\mathcal E_{\mu}$. We call $\sigma=[\sigma_1,\dots,\sigma_k]$ the \emph{shape sequence} of $\mathcal E_{\mu}$. Further, consider the sequence $S=[s_1,\dots,s_x]$ obtained from $\sigma$ by identifying consecutive identical shape descriptions. We call $S$ the \emph{contracted shape sequence} of $\mathcal E_{\mu}$; see Fig.~\ref{fig:preliminaries}.

%%%%%%%%%%%%%%%%%%%%%%%%%
%%%%%%%%%%%%%%%%%%%%%%%%%
%%%%%%%%%%%%%%%%%%%%%%%%%
%%%%%%%%%%%%%%%%%%%%%%%%%
%%%%%%%%%%%%%%%%%%%%%%%%%
%%%%%%%%%%%%%%%%%%%%%%%%%
%%%%%%%%%%%%%%%%%%%%%%%%%
%%%%%%%%%%%%%%%%%%%%%%%%%
%%%%%%%%%%%%%%%%%%%%%%%%%

\section{A Prescribed Edge on the Outer Face}\label{sec:prescribed}
	
%\todo[inline,color=yellow]{Question: Is the concept of special/normal component needed at all?}
%
%Let $\nu$ be a node of $T$ and let $u$ and $v$ be the poles of $\nu$. We say that $G_{\nu}$ is \emph{special} for a pole $w\in \{u,v\}$ if $G_{\nu}$ contains both incoming and outgoing edges incident to $w$. We say that $G_{\nu}$ is \emph{normal} if it is special neither for $u$ nor for $v$. 
%	
%\begin{lemma} \label{le:few-special}
%Let $\mu$ be a P-node of $T$. Suppose that $G_{\mu}$ has an upward planar embedding. Then, for each pole $w$ of $\mu$, there are at most two components of $G_{\mu}$ that are special for $w$.
%\end{lemma}
%
%\begin{proof}
%In every upward planar embedding of $G_{\mu}$, all the edges outgoing a vertex $w$ are consecutive in clockwise order around $w$; that is, the planar embedding of $G_{\mu}$ is \emph{bimodal}, see~\cite{bdl-udtd-94,gt-ccu-01}. However, if (at least) three components of $G_{\mu}$ contain both incoming and outgoing edges of $w$, no planar embedding of $G_{\mu}$ is bimodal.
%\end{proof}

%

Let $G$ be an $n$-vertex biconnected directed partial $2$-tree and $T$ be its SPQ-tree rooted at any Q-node $\rho^*$, which corresponds to an edge $e^*$ of $G$. In this section, we show an algorithm that computes the feasible set $\mathcal F_{\mu}$ of every node $\mu$ of $T$. Let $u$ and $v$ be the poles of $\mu$. Note that $G$ admits an upward planar embedding such that $e^*$ is incident to the outer face if and only if the feasible set of $\rho^*$ is non-empty. Hence, the algorithm could be applied repeatedly (once for each Q-node as the root) to test the upward planarity of $G$; however, in Section~\ref{sec:unrooting} we devise a more efficient way to handle multiple choices for the root of $T$. We first deal with S-nodes, then with P-nodes, and finally with the root of $T$. For Q-nodes, it is easy to show the following lemma.

\begin{lemma}[\cite{cdf-up-22}]\label{lem:Q_node_general}
For a non-root Q-node $\mu$, $\mathcal{F}_\mu$ can be computed in $\bigoh(1)$ time.
%	Let $\mu$ be a non-root Q-node of $T$. The feasible set $\mathcal{F}_\mu$ of $\mu$ can be computed in $\bigoh(1)$ time.
\end{lemma}

%The algorithm assumes that the feasible sets $\mathcal F_{\nu_1},\dots,\mathcal F_{\nu_k}$ of the children $\nu_1,\dots,\nu_k$ of $\mu$ are known. The algorithm runs in $\bigoh(n+|\mathcal F_{\nu_1}|\cdot |\mathcal F_{\nu_2}|)$ time if $\mu$ is an S-node, in $\bigoh(nk)$ time if $\mu$ is a P-node (in this case, the algorithm also computes some auxiliary information, to be described later), and in $\bigoh(n)$ time if $\mu$ is the root of $T$.

%The algorithm allows us to compute in $\bigoh(n^2)$ time the feasible set $\mathcal F_{\mu}$ of every node $\mu$ of $T$. 
%
%*****Q-nodes with details
%\medskip
%\noindent{\bf Q-nodes.} For a Q-node $\mu$ representing an edge $(u,v)$,  $\mathcal{F}_\mu$ contains only the shape description $\shapeDesc{0}{0}{1}{1}{\textrm{out}}{\textrm{out}}{\textrm{in}}{\textrm{in}}$ if $(u,v)$ is outgoing $u$ and only $\shapeDesc{0}{0}{1}{1}{\textrm{in}}{\textrm{in}}{\textrm{out}}{\textrm{out}}$ otherwise. This leads to the following lemma.
%
%\begin{lemma}[\cite{cdf-up-22}]\label{lem:Q_node_general}
%	Let $\mu$ be a non-root Q-node of $T$. The feasible set $\mathcal{F}_\mu$ of $\mu$ can be computed in $\bigoh(1)$ time.
%\end{lemma}

\medskip
\noindent{\bf S-nodes.} We improve an algorithm from~\cite{cdf-up-22}. Let $\nu_1$ and $\nu_2$ be the children of $\mu$ in $T$, let $n^\mu_1=|V(G_{\nu_1})|$ and $n^\mu_2=|V(G_{\nu_2})|$, and let $w$ be the vertex shared by $G_{\nu_1}$ and $G_{\nu_2}$. Furthermore, let $n^\mu_3$ be the number of vertices in the subgraph $H_{\mu}$ of $G$ induced by $V(G)\setminus V(G_\mu) \cup \{u,v\}$. Note that $n^\mu_3=|V(G)|-(n^\mu_1+n^\mu_2)+3$. We distinguish two cases, depending on which of $n^\mu_1$, $n^\mu_2$, and $n^\mu_3$ is largest.

If $n^\mu_3\geq \max(n^\mu_1,n^\mu_2)$, we proceed as in~\cite[Lemma 6]{cdf-up-22}, by combining every shape description in $\mathcal{F}_{\nu_1}$ with every shape description in $\mathcal{F}_{\nu_2}$; for every such combination, the algorithm assigns the angles at $w$ in the outer face with every possible label in $\{-1,0,1\}$. If the combination and assignment result in a shape description $s$ of $\mu$ (the satisfaction of the properties of Theorem~\ref{th:upward-conditions} are checked here), the algorithm adds $s$ to $\mathcal{F}_\mu$. This allows us to compute the feasible set $\mathcal{F}_\mu$ of $\mu$ in time $\bigoh(n+|\mathcal{F}_{\nu_1}|\cdot|\mathcal{F}_{\nu_2}|)$, which is in $\bigoh(n+n^\mu_1\cdot n^\mu_2)$, as $|\mathcal{F}_{\nu_1}|\in \bigoh(n^\mu_1)$ and $|\mathcal{F}_{\nu_2}|\in \bigoh(n^\mu_2)$ by Lemma~\ref{le:shape-generation}.

The most interesting case is when, say, $n^\mu_1\geq \max(n^\mu_2,n^\mu_3)$. Here, in order to keep the overall runtime in $\bigoh(n^2)$, we cannot combine every shape description in $\mathcal{F}_{\nu_1}$ with every shape description in $\mathcal{F}_{\nu_2}$. Rather, we proceed as follows. Note that every shape description in $\mathcal{F}_{\mu}$ whose absolute value of the (left- or right-) turn-number exceeds $n^\mu_3+4$ does not result in an upward planar embedding of $G$, by Property~UP3 of Theorem~\ref{th:upward-conditions} and since the absolute value of the turn-number of any path in any upward planar embedding of $H_{\mu}$ does not exceed $n^\mu_3$. We hence construct an $(n^\mu_3+4)$-universal set $\mathcal S$ in $\bigoh(n^\mu_3)$ time by~Lemma~\ref{le:shape-generation}, and then test whether each shape description $s$ in $\mathcal S$ belongs to the feasible set $\mathcal{F}_{\mu}$ of $\mu$. In order to do that, we consider every shape description $s_2$ in $\mathcal{F}_{\nu_2}$ individually. There are $\bigoh(1)$ shape descriptions in $\mathcal{F}_{\nu_1}$ which combined with $s_2$ might result in $s$, since the turn numbers add to each other when combining the shape descriptions in $\mathcal{F}_{\nu_1}$ and $\mathcal{F}_{\nu_2}$, with a constant offset. Hence, by Lemma~\ref{le:shape-data-structure}, we check in $\bigoh(1)$ time if there is a shape description $s_1$ in $\mathcal{F}_{\nu_1}$ which combined with $s_2$ leads to $s$. The running time of this procedure is hence $\bigoh(n+n^\mu_2\cdot n^\mu_3)$, as $|\mathcal{F}_{\nu_2}|\in \bigoh(n^\mu_2)$ and $|\mathcal{S}|\in \bigoh(n^\mu_3)$ by Lemma~\ref{le:shape-generation}. This yields the following.

% proceed as follows. Note that every shape description of $G_{\mu}$ whose absolute value of the (left- or right-) turn-number exceeds $n^\mu_3+4$ does not result in an upward planar embedding of $G$, by Property~UP3 of Theorem~\ref{th:upward-conditions} and since the absolute value of the (left- or right-) turn-number of any path in any upward planar embedding of $H_{\mu}$ does not exceed $n^\mu_3$. We combine every shape description $s_2$ in $\mathcal{F}_{\nu_2}$ with every shape description $s$ in an $(n^\mu_3+4)$-universal set $\mathcal S$. There are $\bigoh(1)$ shape descriptions in $\mathcal{F}_{\nu_3}$ which combined with $s_2$ might result in $s$, since the turn numbers add to each other when combining the shape descriptions, with a constant offset. Hence, by Lemma~\ref{le:shape-data-structure}, we check in $\bigoh(1)$ time if there is a shape description $s_3$ in $\mathcal{F}_{\nu_3}$ which combined with $s_2$ leads to $s$. This yields the following.

\begin{lemma}\label{lem:S_node_general}
	Let $\mu$ be an S-node of $T$ with children $\nu_1$ and $\nu_2$. Given the feasible sets $\mathcal{F}_{\nu_1}$ and $\mathcal{F}_{\nu_2}$ of $\nu_1$ and $\nu_2$, respectively, the feasible set $\mathcal{F}_\mu$ of $\mu$ can be computed in 
	$\bigoh(n+\min\{n^\mu_1\cdot n^\mu_2,~n^\mu_2\cdot n^\mu_3,~n^\mu_1\cdot n^\mu_3\})$ time.
\end{lemma}

%
%*********
%
%
%In order to compute the feasible set $\mathcal{F}_\mu$ of an S-node $\mu$, we use an algorithm devised in~\cite{cdf-up-22}. Let $\nu_1$ and $\nu_2$ be the children of $\mu$ in $T$ and let $w$ be the vertex shared by $G_{\nu_1}$ and $G_{\nu_2}$. Roughly speaking, the algorithm in~\cite{cdf-up-22} combines every shape description $s_1$ in $\mathcal{F}_{\nu_1}$ with every shape description $s_2$ in $\mathcal{F}_{\nu_2}$; for every such a combination, the algorithm assigns the angles at $w$ in the outer face with every possible label in $\{-1,0,1\}$. Whenever the combination and the assignment result in a shape description $s$ of $\mu$ (the satisfaction of the properties of Theorem~\ref{th:upward-conditions} are checked here), the algorithm adds $s$ to $\mathcal{F}_\mu$. This yields the following lemma.
%
%\begin{lemma}[\cite{cdf-up-22}]\label{lem:S_node_general}
%	Let $\mu$ be an S-node of $T$ with children $\nu_1$ and $\nu_2$. Given the feasible sets $\mathcal{F}_{\nu_1}$ and $\mathcal{F}_{\nu_2}$ of $\nu_1$ and $\nu_2$, respectively, the feasible set $\mathcal{F}_\mu$ of $\mu$ can be computed in $\bigoh(n+|\mathcal{F}_{\nu_1}|\cdot|\mathcal{F}_{\nu_2}|)$ time.
%\end{lemma}

\medskip
\noindent{\bf P-nodes.}  To compute the feasible set $\mathcal F_{\mu}$ of a P-node $\mu$ from the feasible sets $\mathcal F_{\nu_1},\dots,\mathcal F_{\nu_k}$ of its children, the algorithm constructs an $n$-universal set $\mathcal S$ in $\bigoh(n)$ time by Lemma~\ref{le:shape-generation}. Then it examines every shape description $s\in \mathcal S$ and decides whether it belongs to $\mathcal F_{\mu}$. Hence, we focus on a single shape description $s$ and give an algorithm that decides in $\bigoh(k)$ time whether $s$ belongs to $\mathcal F_{\mu}$. 

The basic structural tool we need for our algorithm is the following lemma. We call \emph{generating set} $\mathcal G(s)$ of a shape description $s$ the set of contracted shape sequences that the pertinent graph of any P-node with poles $u$ and $v$ can have in a $uv$-external upward planar embedding with shape description $s$. 

%An algorithm with the same objective and running time was presented in~\cite{cdf-up-22}. The algorithm we present here improves upon the one in~\cite{cdf-up-22} in three ways. First, it works for a directed partial $2$-tree that is not necessarily expanded; as explained in the introduction, the expansion of a directed partial $2$-tree might result into a directed graph which is not a partial $2$-tree. Second, it is much simpler. Third, it computes some auxiliary information that we are going to need in Section~\ref{sec:unrooting} in order to devise an $\bigoh(n^2)$-time upward planarity testing algorithm. 

\begin{lemma}[$\star$]\label{le:shape-descriptions-structure}
For any shape description $s$, $\mathcal G(s)$ has size $\bigoh(1)$ and can be constructed in $\bigoh(1)$ time. Also, any sequence in $\mathcal G(s)$ has~$\bigoh(1)$~length.
\end{lemma}

A contracted shape sequence $S\in \mathcal G(s)$ is \emph{realizable} by $\mu$ if there exists a $uv$-external upward planar embedding of $G_{\mu}$ whose contracted shape sequence is a subsequence of $S$ containing the first and last elements of $S$. 

We now describe an algorithm that decides in $O(k)$ time whether $s$ belongs to $\mathcal F_{\mu}$. Also, for each contracted shape sequence $S=[s_1,\dots,s_x]$ in the generating set $\mathcal G(s)$ of $s$, the algorithm computes and stores the following information: 

\begin{itemize}
	\item Three labels $\textsc{f}_1(\mu,S)$, $\textsc{f}_2(\mu,S)$, and $\textsc{f}_3(\mu,S)$ which reference three distinct children $\nu_i$ of $\mu$ such that $s_1\in \mathcal F_{\nu_i}$. 
	\item Three labels $\textsc{l}_1(\mu,S)$, $\textsc{l}_2(\mu,S)$, and $\textsc{l}_3(\mu,S)$ which reference three distinct children $\nu_i$ of $\mu$ such that $s_x\in \mathcal F_{\nu_i}$.
	\item Two labels $\textsc{uf}_1(\mu,S)$ and $\textsc{uf}_2(\mu,S)$ which reference two distinct children $\nu_i$ of $\mu$ such that $\mathcal F_{\nu_i}$ does not contain any shape description in $S$. 
\end{itemize}
For each label type, if the number of children with the described properties is smaller than the number of labels, then labels with larger indices are $\nullo$. We call \emph{the set of relevant labels for $\mu$ and $S$} the set of labels described above. 

The algorithm is as follows. First, by Lemma~\ref{le:shape-descriptions-structure}, we construct $\mathcal G(s)$ in $\bigoh(1)$ time. Then we consider each sequence $S=[s_1,\dots,s_x]$ in $\mathcal G(s)$. By Lemma~\ref{le:shape-descriptions-structure}, there are $\bigoh(1)$ such sequences, each with length $\bigoh(1)$. We decide whether $S$ is realizable by $\mu$ and compute the set of relevant labels for $\mu$ and $S$ as follows. 

We initialize all the labels to $\nullo$ and process $\nu_1,\dots,\nu_k$ one by one. For each $\nu_i$, by Lemma~\ref{le:shape-data-structure} we test in $\bigoh(1)$ time which of the shape descriptions $s_1,\dots,s_x$ belong to $\mathcal F_{\nu_i}$ and update the labels accordingly. For example, if $s_1\in\mathcal F_{\nu_i}$, then we update $\textsc{f}_j(\mu,S)=\nu_i$ for the smallest  $j\in\{1,2,3\}$ with $\textsc{f}_j(\mu,S)=\nullo$. 

After processing $\nu_1,\dots,\nu_k$, we decide whether $S$ is realizable by $\mu$ as follows. If $\textsc{uf}_1(\mu,S)\neq \nullo$, then $S$ is not realizable by $\mu$. Otherwise, each feasible set $\mathcal F_{\nu_i}$ contains a shape description among $s_1,\dots,s_x$. Still, we have to check whether $\mathcal F_{\nu_i}$ contains $s_1$ and $\mathcal F_{\nu_j}$ contains $s_x$, for two distinct nodes $\nu_i$ and $\nu_j$. If $\textsc{f}_1(\mu,S)=\nullo$ or $\textsc{l}_1(\mu,S)=\nullo$, then $S$ is not realizable by $\mu$, as the feasible set of no child contains $s_1$ or $s_x$, respectively. Otherwise, if $\textsc{f}_1(\mu,S)\neq \textsc{l}_1(\mu,S)$, then $S$ is realizable by $\mu$, as $\textsc{f}_1(\mu,S)$ can be assigned with $s_1$ and $\textsc{l}_1(\mu,S)$ with $s_x$.  Otherwise, if $\textsc{f}_2(\mu,S)\neq \nullo$ or $\textsc{l}_2(\mu,S)\neq \nullo$, then $S$ is realizable by $\mu$, as $\textsc{f}_2(\mu,S)$ can be assigned with $s_1$ and $\textsc{l}_1(\mu,S)$ with $s_x$, or $\textsc{f}_1(\mu,S)$ can be assigned with $s_1$ and $\textsc{l}_2(\mu,S)$ with $s_x$, respectively. Otherwise, $S$ is not realizable by $\mu$, as $s_1$ and $s_x$ are in the feasible set of a single child $\textsc{f}_1(\mu,S)=\textsc{l}_1(\mu,S)$~of~$\mu$.

% Note that the labels $\textsc{f}_3(\mu,S)$, $\textsc{l}_3(\mu,S)$, and $\textsc{uf}_2(\mu,S)$ are not used by the described algorithm, however they will be needed in the algorithm presented in Section~\ref{sec:unrooting}.

Finally, we have that $s$ belongs to $\mathcal F_{\mu}$ if and only if there exists a contracted shape sequence $S$ in the generating set $\mathcal G(s)$ of $s$ which is realizable by $\mu$.

%In order to check that, we construct an $\bigoh(1)$-size bipartite graph $\mathcal A(\mu,S)$ in which one node family contains two nodes labeled $s_1$ and $s_x$, and the other node family contains a node for each child $\nu_i$ of $\mu$ which appears in the set $\{\textsc{f}_1(\mu,S), \textsc{f}_2(\mu,S)\}\cup \{\textsc{l}_1(\mu,S), \textsc{l}_2(\mu,S)\}$. Then $\mathcal F_{\nu_i}$ contains $s_1$ and $\mathcal F_{\nu_j}$ contains $s_x$, for two distinct children $\nu_i$ and $\nu_j$ of $\mu$, if and only if  

%if $\textsc{f}_1(\mu,S)=\textsc{ext}_1(\mu,S)=\nullo$, or if $\textsc{l}_1(\mu,S)=\textsc{ext}_1(\mu,S)=\nullo$, or if $\textsc{f}_1(\mu,S)=\textsc{l}_1(\mu,S)=\textsc{ext}_2(\mu,S)=\nullo$, then $S$ is not realizable by $\mu$. Otherwise, $S$ is realizable by $\mu$.

\begin{lemma}[$\star$]\label{lem:P_node_general}
	Let $\mu$ be an P-node of $T$ with children $\nu_1,\dots,\nu_k$. Given their feasible sets $\mathcal{F}_{\nu_1},\dots,\mathcal{F}_{\nu_k}$, the feasible set $\mathcal{F}_\mu$ of $\mu$ can be computed in $\bigoh(nk)$ time. Further, for every shape description $s$ in an $n$-universal set $\mathcal S$ and every contracted shape sequence $S$ in the generating set $\mathcal G(s)$ of $s$, the set of relevant labels for $\mu$ and $S$ can be computed and stored in overall $\bigoh(nk)$ time and space. 
\end{lemma}

\medskip
\noindent{\bf Root.} As in~\cite{cdf-up-22}, the root $\rho^*$ of $T$ is treated as a P-node with two children, whose pertinent graphs are $e^*$ and the pertinent graph of the child $\sigma^*$ of $\rho^*$ in~$T$.

\begin{lemma}[\cite{cdf-up-22}]\label{lem:root}
Given the feasible set $\mathcal{F}_{\sigma^*}$, the feasible set $\mathcal{F}_{\rho^*}$ of the root $\rho^*$ of $T$ can be computed in $\bigoh(n)$ time.
\end{lemma}

\section{No Prescribed Edge on the Outer Face}\label{sec:unrooting}

In this section, we show an $\bigoh(n^2)$-time algorithm to test the upward planarity of a biconnected directed partial $2$-tree $G$.
Let $e_1,\dots,e_m$ be any order of the edges of $G$. For $i=1,\dots,m$, let $\rho_i$ be the Q-node of the SPQ-tree $T$ of $G$ corresponding to $e_i$ and $T_i$ be the rooted tree obtained by selecting $\rho_i$ as the root of $T$. For a node $\mu$ of $T$, distinct choices for the root of $T$ define different pertinent graphs $G_{\mu}$ of $\mu$. Thus, we change the previous notation and denote by $G_{\mu\rightarrow \tau}$ and $\mathcal F_{\mu\rightarrow \tau}$ the pertinent graph and the feasible set of a node $\mu$ when its parent is a node $\tau$. We denote by $\mathcal F_{\rho_i}$ the feasible set of the root $\rho_i$ of~$T_i$.

Our algorithm performs traversals of $T_1,\dots,T_m$. The traversal of $T_1$ is special; it is a bottom-up traversal using the results from Section~\ref{sec:prescribed} to compute the feasible set $\mathcal F_{\mu\rightarrow \tau}$ of every node $\mu$ with parent $\tau$ in $T_1$, as well as auxiliary information that is going to be used by later traversals. For $i=2,\dots,m$, we perform a top-down traversal of $T_i$ that computes the feasible set $\mathcal F_{\mu\rightarrow \tau}$ of every node $\mu$ with parent $\tau$ in $T_i$. Due to the information computed by the traversal of $T_1$, this can be carried out in $\bigoh(n)$ time for each P-node.  Further, the traversal of $T_i$ visits a subtree of $T_i$ only if that has not been visited ``in the same direction'' during a traversal $T_j$ with $j<i$. We start with two auxiliary lemmas. 

\begin{lemma}[$\star$]\label{le:empty-climbs}
	Suppose that, for some $i\in \{1,\dots,m\}$, a node $\mu$ with parent $\tau$ has a child $\nu_j$ in $T_i$ such that $\mathcal F_{\nu_j\rightarrow\mu}=\emptyset$. Then $\mathcal F_{\mu\rightarrow\tau}=\emptyset$.
\end{lemma}

\begin{lemma}[$\star$]\label{le:two-bad-children}
Suppose that a node $\mu$ has two neighbors $\nu_j$ and $\nu_k$ such that $\mathcal F_{\nu_j\rightarrow\mu}=\mathcal F_{\nu_k\rightarrow\mu}=\emptyset$. Then $G$ admits no upward planar embedding.
\end{lemma}

\medskip
\noindent{\bf Bottom-up Traversal of $\mathbf T_1$.}
The first step of the algorithm consists of a bottom-up traversal of $T_1$. This step either rejects the instance (i.e., it concludes that $G$ admits no upward planar embedding) or computes and stores, for each non-root node $\mu$ of $T_1$ with parent $\tau$, the feasible set $\mathcal F_{\mu\rightarrow \tau}$ of $\mu$, as well as the feasible set $\mathcal F_{\rho_1}$ of the root $\rho_1$. Further, if $\mu$ is an S- or P-node, it also computes the following information.
\begin{itemize}
	\item A label $\textsc{p}(\mu)$ referencing the parent $\tau$ of $\mu$ in $T_1$.
	\item A label $\textsc{uc}(\mu)$ referencing a node $\nu$ such that $\mathcal F_{\nu\rightarrow \mu}$ has not been computed. Initially this is $\tau$, and once $\mathcal F_{\tau\rightarrow \mu}$ is computed, this label changes to $\nullo$. 
	\item A label $\textsc{b}(\mu)$ referencing any neighbor $\nu$ of $\mu$ such that $\mathcal F_{\nu\rightarrow \mu}=\emptyset$. This label remains $\nullo$ until such neighbor is found.
\end{itemize}

Finally, if $\mu$ is a P-node, for each shape description $s$ in an $n$-universal set $\mathcal S$ and each contracted shape sequence $S=[s_1,\dots,s_x]$ in the generating set $\mathcal G(s)$ of $s$, the algorithm computes and stores the set of relevant labels for $\mu$ and~$S$.

The bottom-up traversal of $T_1$ computes the feasible set $\mathcal F_{\mu\rightarrow \tau}$ in $\bigoh(1)$ time by Lemma~\ref{lem:Q_node_general}, for any Q-node $\mu\neq \rho_1$ with parent $\tau$. When an S- or P-node $\mu$ with parent $\tau$ is visited, the algorithm stores in $\textsc{p}(\mu)$ and $\textsc{uc}(\mu)$ a reference to $\tau$. Then it considers $\textsc{b}(\mu)$. Suppose that $\textsc{b}(\mu)\neq \nullo$ (the label $\textsc{b}(\mu)$ might have been assigned a value different from $\nullo$ when visiting a child of $\mu$). By Lemma~\ref{le:empty-climbs} we have $\mathcal F_{\mu\rightarrow \tau}=\emptyset$, hence if $\textsc{b}(\tau)\neq \nullo$, then by Lemma~\ref{le:two-bad-children}, the algorithm rejects the instance, otherwise it sets $\textsc{b}(\tau)=\mu$ and concludes the visit of $\mu$. 
Suppose next that $\textsc{b}(\mu)=\nullo$. Then we have $\mathcal F_{\nu_j\rightarrow \mu}\neq \emptyset$, for every child $\nu_j$ of $\mu$, thus $\mathcal F_{\mu\rightarrow \tau}$ is computed using Lemma~\ref{lem:S_node_general} or~\ref{lem:P_node_general}, if $\mu$ is an S-node or a P-node, respectively. If  $\mathcal F_{\mu\rightarrow \tau}=\emptyset$, then the algorithm checks whether $\textsc{b}(\tau)\neq \nullo$ (and then it rejects the instance) or not (and then it sets $\textsc{b}(\tau)=\mu$). This concludes the visit of $\mu$.
Finally, when the algorithm reaches $\rho_1$, it checks whether $\textsc{b}(\rho_1)=\nullo$ and if the test is positive, then it concludes that $\mathcal F_{\rho_1}=\emptyset$. Otherwise, it computes $\mathcal F_{\rho_1}$ by means of Lemma~\ref{lem:root} and completes the traversal of $T_1$.

\medskip
\noindent{\bf Top-Down Traversal of $\mathbf T_i$.} The top-down traversal of $T_i$ computes $\mathcal F_{\mu\rightarrow \tau}$, for each non-root node $\mu$ with parent $\tau$ in $T_i$, as well as $\mathcal F_{\rho_i}$. For each S- or P-node $\mu$, the labels $\textsc{uc}(\mu)$ and $\textsc{b}(\mu)$ might be updated during the traversal of $T_i$, while $\textsc{p}(\mu)$ and the sets of relevant labels are never altered after the traversal of $T_1$. The traversal of $T_i$ visits a node $\mu$ with parent $\tau$ only if $\mathcal F_{\mu\rightarrow \tau}$ has not been computed yet; this information is retrieved in $\bigoh(1)$ time from the label $\textsc{uc}(\tau)$.

%. By Lemmata~\ref{le:two-bad-children} and~\ref{le:root-is-good}, this either allow the algorithm to conclude that $G$ admits no upward planar embedding, or the algorithm proceeds with the traversal of $T_{i+1}$. If no top-down traversal concludes that $G$ admits an upward planar embedding, then again by Lemma~\ref{le:root-is-good}, the algorithm can conclude that $G$ admits no upward planar embedding.

%For any Q-node $\mu\neq \rho_i$, the feasible set $\mathcal F_{\mu\rightarrow \tau}$ is computed in $\bigoh(1)$ time by Lemma~\ref{lem:Q_node_general}; actually, $\mathcal F_{\mu\rightarrow \tau}$ has been already computed in the bottom-up traversal of $T_1$, except for $\mu=\rho_1$, for which $\mathcal F_{\mu\rightarrow \tau}$ is computed in the top-down traversal of $T_2$. 

When the traversal visits an S- or P-node $\mu$ with parent $\tau$ and children $\nu_1,\dots,\nu_k$, it proceeds as follows. Note that $\textsc{p}(\mu)\neq \tau$, as otherwise $\mathcal F_{\mu\rightarrow \tau}$ would have been already computed. Then we have  $\textsc{p}(\mu)=\nu_{j^*}$, for some $j^*\in\{1,\dots,k\}$. 

If $\textsc{uc}(\mu)=\nu_{j^*}$, then before computing $\mathcal F_{\mu\rightarrow \tau}$, the algorithm descends in $\nu_{j^*}$ in order to compute $\mathcal F_{\nu_{j^*}\rightarrow \mu}$. Otherwise, $\mathcal F_{\nu_j\rightarrow \mu}$ has been computed for $j=1,\dots,k$. 

If $\textsc{b}(\mu)=\nu_{j}$, for some $j\in\{1,\dots,k\}$, then by Lemma~\ref{le:empty-climbs} we have $\mathcal F_{\mu\rightarrow \tau}=\emptyset$, hence if $\textsc{b}(\tau)\neq \nullo$ and $\textsc{b}(\tau)\neq \mu$, then the algorithm rejects the instance by Lemma~\ref{le:two-bad-children}, otherwise it sets $\textsc{b}(\tau)= \mu$ and concludes the visit of $\mu$. 
Conversely, if $\textsc{b}(\mu)=\nullo$ or $\textsc{b}(\mu)=\tau$, then $\mathcal F_{\nu_j\rightarrow \mu}\neq \emptyset$ for $j=1,\dots,k$. The algorithm then computes $\mathcal F_{\mu\rightarrow \tau}$, as described below. Afterwards, if $\textsc{uc}(\tau)= \mu$, the algorithm sets $\textsc{uc}(\tau)= \nullo$. Further, if $\mathcal F_{\mu\rightarrow \tau}=\emptyset$, the algorithm checks whether $\textsc{b}(\tau)\neq \nullo$ (and then rejects the instance) or not (and then sets $\textsc{b}(\tau)=\mu$). 

The computation of $\mathcal F_{\mu\rightarrow \tau}$ distinguishes the case when $\mu$ is an S-node or a P-node. If $\mu$ is an S-node, then the computation of $\mathcal F_{\mu\rightarrow \tau}$ is done by means of Lemma~\ref{lem:S_node_general}. The running time of the procedure for the S-nodes sums up to $\bigoh(n^2)$, over all S-nodes and all traversals of $T$. If $\mu$ is a P-node, then the computation of $\mathcal F_{\mu\rightarrow \tau}$ cannot be done by just applying the algorithm from Lemma~\ref{lem:P_node_general}, as that would take $\Theta(n^3)$ time for all P-nodes and all traversals of $T$. Instead, the information computed when traversing $T_1$ allows us to determine in $\bigoh(1)$ time whether any shape description is in $\mathcal F_{\mu\rightarrow \tau}$. This results in an $\bigoh(n)$ time for processing $\mu$ in $T_i$, which sums up to $\bigoh(nk)$ time over all traversals of $T$, and thus in a $\bigoh(n^2)$ total running time for the entire algorithm.

%If $\textsc{b}(\mu)\neq \nullo$ and $\textsc{b}(\mu)\neq \tau$, then $\mu$ has a child $\nu_j$ (the one referenced by $\textsc{b}$) such that $\mathcal F_{\nu_j\rightarrow \mu}=\emptyset$. We handle this case as follows. By Lemma~\ref{le:empty-climbs}, we set $\mathcal F_{\mu\rightarrow \tau}=\emptyset$. We then check whether the realization that $\mathcal F_{\mu\rightarrow \tau}=\emptyset$ allows us to conclude that $G$ admits no upward planar embedding. Indeed, if $\textsc{b}(\tau)\neq \nullo$ and $\textsc{b}(\tau)\neq \mu$, then by Lemma~\ref{le:two-bad-children} we conclude that $G$ admits no upward planar embedding; otherwise, we set $\textsc{b}(\tau)= \mu$ and continue the traversal of $T_1$. Note that all these checks and modifications can be performed in $\bigoh(1)$ time. f $\textsc{b}(\mu)= \nullo$ or $\textsc{b}(\mu)= \tau$, 

The algorithm determines $\mathcal F_{\mu\rightarrow \tau}$ by examining each shape description $s$ in an $n$-universal set $\mathcal S$, which has $\bigoh(n)$ elements and is constructed in $\bigoh(n)$ time by Lemma~\ref{le:shape-generation}, and deciding whether it is in $\mathcal F_{\mu\rightarrow \tau}$ or not. This is done as follows. We construct in $\bigoh(1)$ time the generating set $\mathcal G(s)$ of $s$, by Lemma~\ref{le:shape-descriptions-structure}. Recall that $\mathcal G(s)$ contains $\bigoh(1)$ contracted shape sequences, each with length $\bigoh(1)$. For each sequence $S=[s_1,\dots,s_x]$ in $\mathcal G(s)$, we test whether $S$ is realizable by $\mu$ as follows. 

\begin{itemize}
	\item If $\textsc{uf}_2(\mu,S)\neq \nullo$, or if $\textsc{uf}_1(\mu,S)\neq \nullo$ and $\textsc{uf}_1(\mu,S)\neq \tau$, then there exists a child $\nu_j$ of $\mu$ in $T_i$ such that $\mathcal F_{\nu_j\rightarrow \mu}$ does not contain any shape description in $S$. Then we conclude that $S$ is not realizable by $\mu$.
	\item Otherwise, we test whether $\mathcal F_{\nu_{j^*}\rightarrow \mu}$ contains any shape description among the ones in $S$. If not, $S$ is not realizable by $\mu$. Otherwise, for $j=1,\dots,k$, $\mathcal F_{\nu_j\rightarrow \mu}$ contains a shape description in $S$. However, this does not imply that $S$ is realizable by $\mu$, as we need to ensure that $s_1\in  \mathcal F_{\nu_j\rightarrow \mu}$ and $s_x\in  \mathcal F_{\nu_l\rightarrow \mu}$ for two distinct children $\nu_j$ and $\nu_l$ of $\mu$ in $T_i$. This can be tested as follows. 
	
	We construct a bipartite graph $\mathcal B_{\mu \rightarrow \tau}(S)$ in which one family has two vertices labeled $s_1$ and $s_x$. The other one has a vertex for each child of $\mu$ in the set $\{\textsc{f}_1(\mu,S),\textsc{f}_2(\mu,S),\textsc{f}_3(\mu,S),$ $\textsc{l}_1(\mu,S),$ $\textsc{l}_2(\mu,S),\textsc{l}_3(\mu,S),\nu_{j^*}\}$. The graph $\mathcal B_{\mu \rightarrow \tau}(S)$ contains an edge between the vertex representing a child $\nu_j$ of $\mu$ and a vertex representing $s_1$ or $s_x$ if $s_1$ or $s_x$ belongs to $\mathcal F_{\nu_j\rightarrow \mu}$, respectively. We now have that $s_1\in  \mathcal F_{\nu_j\rightarrow \mu}$ and $s_x\in  \mathcal F_{\nu_l\rightarrow \mu}$ for two distinct children $\nu_j$ and $\nu_l$ of $\mu$ in $T_i$ (and thus $S$ is realizable by $\mu$) if and only if $\mathcal B_{\mu \rightarrow \tau}(S)$ contains a size-$2$ matching, which can be tested in $\bigoh(1)$ time. 
\end{itemize}

Testing whether $S$ is realizable by $\mu$ can be done in $\bigoh(1)$ time, as it only requires to check $\bigoh(1)$ labels, to find a size-$2$ matching in a $\bigoh(1)$-size graph, and to check $\bigoh(1)$ times whether a shape description belongs to a feasible set. The last operation requires $\bigoh(1)$ time by Lemma~\ref{le:shape-data-structure}. We conclude that $s$ is in $\mathcal F_{\mu\rightarrow \tau}$ if and only if at least one contracted shape sequence $S$ in $\mathcal G(s)$ is realizable by $\mu$. This concludes the description of how the algorithm handles a P-node.

Finally, $\mathcal F_{\rho_i}$ is computed in $\bigoh(n)$ time by Lemma~\ref{lem:root}. We get the following.

\begin{lemma}[$\star$]\label{le:bico-summing-up}
The described algorithm runs in $\bigoh(n^2)$ time and either correctly concludes that $G$ admits no upward planar embedding, or computes the feasible sets $\mathcal F_{\rho_1},\dots,\mathcal F_{\rho_m}$.
\end{lemma}

\section{Single-Connected Graphs}\label{se:simply-connected}

%\todo[inline]{Currently 70 lines; 3 lines are todos, maybe some 10 lines can be saved by squeezing text between theorems and proof sketches}

In this section, we extend Lemma~\ref{le:bico-summing-up} from the biconnected case to arbitrary partial $2$-trees. To this end, we obtain a general lemma that allows us to test upward planarity of digraphs from the feasible sets of biconnected components.

\begin{lemma}[$\star$]
    Let $G$ be an $n$-vertex digraph. Let $B_1$, \dots, $B_t$ be the maximal biconnected components of $G$. For $i \in [t]$, let the edges of $B_i$ be $e^i_1$, \dots, $e^i_{m_i}$, and the respective Q-nodes in the SPQR-tree of $B_i$ be $\rho^i_1$, \dots, $\rho^i_{m_i}$. There is an algorithm that, given $G$ and the feasible sets $\mathcal{F}_{\rho^i_j}$ for each $i \in [t]$ and $j \in [m_i]$, in time $\bigoh(n^2)$ correctly decides whether $G$ admits an upward planar embedding.
    \label{lem:from_bi_to_single}
\end{lemma}

Note that Lemma~\ref{lem:from_bi_to_single} holds for all digraphs, not only partial $2$-trees. In fact, it generalizes~\cite[Section 5]{cdf-up-22},
%corr/abs-2203.05364
where an analogous statement has been shown for all expanded graphs. Our main result follows from Lemmas~\ref{lem:from_bi_to_single} and~\ref{le:bico-summing-up}. 
%The result of~\cite{cdf-up-22} is however insufficient for our purposes, as partial $2$-trees are not closed under expansion.

\begin{theorem}[$\star$]
 \label{thm:main}
        Let $G$ be an $n$-vertex directed partial $2$-tree. It is possible to determine whether $G$ admits an upward planar embedding in time $\bigoh(n^2)$.
%Furthermore, if $\tau$ has the distribution property, it is possible to determine whether $G$ admits an upward planar drawing in time $\bigoh(n (\alpha(G)+\tau\cdot n))$.
\end{theorem}
%\ifshort
%%Possible proof sketch
%\begin{proof}[Sketch]
%    If $G$ is biconnected, run Lemma~\ref{le:bico-summing-up} on $G$. Otherwise, compute the blocks $B_1$, \dots, $B_t$ of $G$. Run Lemma~\ref{le:bico-summing-up} on each block. If some block does not admit an upward planar embedding, we can immediately conclude that $G$ does not admit one as well. Otherwise, invoke Lemma~\ref{lem:from_bi_to_single} with the obtained feasible sets. As for the running time, the time spent on all invocation of Lemma~\ref{le:bico-summing-up} is upper-bounded by $\bigoh(n^2)$ and invoking Lemma~\ref{lem:from_bi_to_single} also takes time $\bigoh(n^2)$.
%\end{proof}
%\fi

Hence, all that remains now is to prove Lemma~\ref{lem:from_bi_to_single}. To give an intuition of the proof, we start by guessing the root of the block-cut tree of $G$, which corresponds to a biconnected component that is assumed to see the outer face in the desired upward planar embedding of $G$. 
The core of the proof is the following lemma, which states that leaf components can be disregarded as long as certain simple conditions on their parent cut-vertex are met.
%which states that leaf components can always be added on top of an upward planar embedding of the remaining digraph, up to simple conditions on the cut-vertex.

\begin{lemma}[$\star$]
    Consider a rooted block-cut tree of a digraph $G$, its cut vertex $v$ that is adjacent to leaf blocks $B_1$,...,$B_\ell$, and the parent block $P$. Denote by $G^P$ the subgraph $G\left[\left(V(G) \setminus \bigcup_{i\in [\ell]} B_i\right)\cup \{v\}\right]$.
Any upward planar embedding of $G^P$ in which the root block is adjacent to the outer face, can be extended to an embedding of $G$ with the same property if the following conditions hold:
\begin{enumerate}
    \item Each $B_i$ has an upward planar embedding with $v$ on the outer face $f_i$.
    \item If $v$ is a non-switch vertex in $P$, each $B_i$ has an upward planar embedding with $v$ on $f_i$ where the angle at $v$ in $f_i$ is not small.
    \item If there is $j \in [\ell]$ such that $v$ is a non-switch vertex in $B_j$, and all upward planar embeddings of $B_j$ with $v$ on $f_j$ have a small angle at $v$ in $f_j$, then for all $i \in [\ell]$ s.t. $i \ne j$ and $v$ is a non-switch vertex in $B_i$, $B_i$ has an upward planar embedding with $v$ on $f_i$ where the angle at $v$ in $f_i$ is flat.
\end{enumerate}
Moreover, if $G$ admits an upward planar embedding in which the root block is adjacent to the outer face, the conditions above are necessarily satisfied.
    \label{lem:children_block}
\end{lemma}

\begin{figure}[htb]
	\centering
	\subfigure[]{\label{fig:single_connected_example_1}\includegraphics[width=.18\textwidth, page=6]{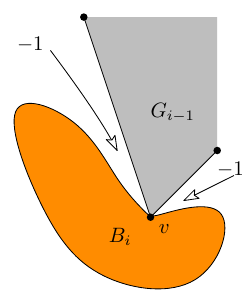}}
	\hfil
	\subfigure[]{\label{fig:single_connected_example_2}\includegraphics[width=.18\textwidth, page=5]{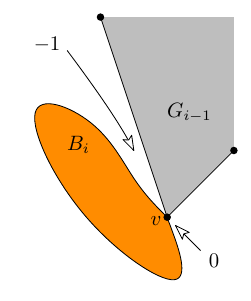}}
	\hfil
	\subfigure[]{\label{fig:single_connected_example_3}\includegraphics[width=.18\textwidth, page=2]{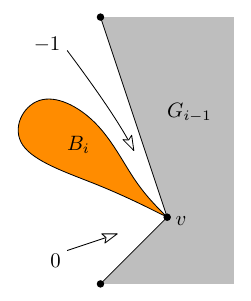}}
	\hfil
	\subfigure[]{\label{fig:single_connected_example_4}\includegraphics[width=.18\textwidth, page=4]{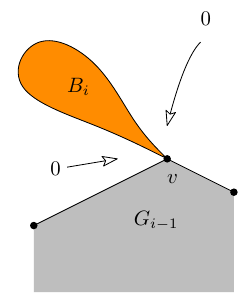}}
	\hfil
	\subfigure[]{\label{fig:single_connected_example_5}\includegraphics[width=.18\textwidth, page=3]{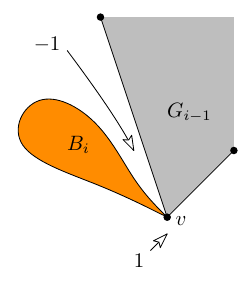}}
	\caption{\label{fig:single_connected_example}Illustrations for the proof of Lemma~\ref{lem:children_block}.		
		%\todo[inline]{Add a drawing similar to (b) but where $B_i$ is inside a flat angle of $G_{i - 1}$}
	}
\end{figure}

%As the conditions of Lemma~\ref{lem:children_block} turn out to be necessary as well, our algorithm simply proceeds upwards along the block-cut tree, checking every time the conditions of the lemma and removing the respective leaf components. If at some point the conditions are not satisfied, we conclude that $G$ is a no-instance; otherwise, any upward planar embedding of the root block can be iteratively extended to an embedding of the whole graph. Finally, finding upward planar embeddings of the blocks is performed via Lemma~\ref{le:bico-summing-up}; the resulting set of feasible shape descriptions is sufficient to verify conditions of Lemma~\ref{lem:children_block}. This concludes the intuitive description of our approach; we now move to formal proofs of Lemma~\ref{lem:children_block} and then Lemma~\ref{lem:from_bi_to_single}.
The proof of Lemma~\ref{lem:children_block} essentially boils down to a case distinction on how the leaf blocks are attached; the cases that need to be considered are intuitively illustrated in Figure~\ref{fig:single_connected_example}. 
With this, we finally have all the components necessary to prove Theorem~\ref{thm:main}. Intuitively, the algorithm proceeds in a leaf-to-root fashion along the block-cut tree, and at each point it checks whether the conditions of Lemma~\ref{lem:children_block} are satisfied. If they are, the algorithm removes the respective leaf components and proceeds upwards, while otherwise we reject the instance.
%. If, on the other hand, the conditions are not satisfied at some point, we conclude that $G$ is a no-instance since by Lemma~\ref{lem:children_block} these conditions are necessary. 
%Finally, finding upward planar embeddings of the blocks is performed via Lemma~\ref{le:bico-summing-up}; the resulting set of feasible shape descriptions is sufficient to verify the conditions of Lemma~\ref{lem:children_block}.
%, and also allows us to construct an upward planar drawing as a witness.
%\todo{See if the previous sections speak about construcing drawings}

%
%\ifshort
%\begin{proof}[Proof Sketch of Lemma~\ref{lem:children_block}]
%    Fix an upward planar embedding $\mathcal{E}^P$ of $G^P$, and upward planar embeddings $\mathcal{E}^{B_1}$, \dots, $\mathcal{E}^{B_\ell}$ of $B_1$, \dots, $B_\ell$ subject to the conditions of the lemma.
%    We start with the embedding $\mathcal{E}^P$ and ``attach'' the embeddings of $B_1$, \dots, $B_\ell$ one by one. On the $i$-th step, we construct an upward planar embedding $\mathcal{E}_i$ of $G_i = G^P \bigcup_{j = 1}^i B_i$ from the embeddings $\mathcal{E}_{i - 1}$  of $G_{i - 1}$ and $\mathcal{E}^{B_i}$ of $B_i$, by ``fitting'' the embedding of $B_i$ inside the face that contains a large or flat angle at $v$ in $\mathcal{E}_{i - 1}$. For that, we make sure that the edges of $B_i$ appear consecutively around $v$, and that the two newly formed angles receive appropriate values to satisfy the conditions of Theorem~\ref{th:upward-conditions}; everything else is directly induced by $\mathcal{E}_{i - 1}$ and $\mathcal{E}^{B_i}$.
%    Figure~\ref{fig:single_connected_example} illustrates the possible cases depending on angle values around $v$.
%\end{proof}
%\fi

\section{Concluding Remarks}\label{sec:conclusions}

We have provided an $\bigoh(n^2)$-time algorithm for testing the upward planarity of $n$-vertex directed partial $2$-trees, substantially improving on the state of the art~\cite{DBLP:journals/siamdm/DidimoGL09}. There are several major obstacles to overcome for improving this runtime to linear; hence, it would be worth investigating whether the quadratic bound is tight. Another interesting direction for future work is to see whether our new techniques can be used to obtain quadratic algorithms for related problems, such as computing orthogonal drawings with the minimum number of bends~\cite{blv-sod-98,dlm-opt-22}.

	\bibliography{biblioSP}
\end{document}